\documentclass[prd,twocolumn,amsmath,amssymb,axodraw]{revtex4}
\usepackage{graphicx}
\begin{document}

\title{\boldmath
Direct Measurements of the Branching Fractions for
$D^0 \to K^-e ^+\nu _e$ and $D^0 \to \pi^-e^+\nu_e$ and Determinations of
the Form Factors $f_{+}^{K}(0)$ and $f^{\pi}_{+}(0)$}
\author{
\begin{small}
M.~Ablikim$^{1}$, J.~Z.~Bai$^{1}$, Y.~Ban$^{10}$, 
J.~G.~Bian$^{1}$, X.~Cai$^{1}$, J.~F.~Chang$^{1}$, 
H.~F.~Chen$^{15}$, H.~S.~Chen$^{1}$, H.~X.~Chen$^{1}$, 
J.~C.~Chen$^{1}$
% ...
\end{small}
\\(BES Collaboration)\\ 
}
\vspace{0.2cm}
\affiliation{
\begin{minipage}{145mm}
$^{1}$ Institute of High Energy Physics, Beijing 100039, People's Republic of China\\
$^{2}$ China Center for Advanced Science and Technology
, 
Beijing 100080, People's Republic of China\\
$^{3}$ Guangxi Normal University, Guilin 541004, People's Republic of China\\
$^{4}$ Henan Normal University, Xinxiang 453002, People's Republic of China\\
$^{5}$ Huazhong Normal University, Wuhan 430079, People's Republic of China\\
$^{6}$ Hunan University, Changsha 410082, People's Republic of China\\
$^{7}$ Liaoning University, Shenyang 110036, People's Republic of China\\
$^{8}$ Nanjing Normal University, Nanjing 210097, People's Republic of China\\
$^{9}$ Nankai University, Tianjin 300071, People's Republic of China\\
$^{10}$ Peking University, Beijing 100871, People's Republic of China\\
$^{11}$ Shandong University, Jinan 250100, People's Republic of China\\
$^{12}$ Shanghai Jiaotong University, Shanghai 200030, People's Republic of China\\
$^{13}$ Sichuan University, Chengdu 610064, People's Republic of China\\
$^{14}$ Tsinghua University, Beijing 100084, People's Republic of China\\
$^{15}$ University of Science and Technology of China, Hefei 230026, 
    People's Republic of China\\
$^{16}$ Wuhan University, Wuhan 430072, People's Republic of China\\
$^{17}$ Zhejiang University, Hangzhou 310028, People's Republic of China\\
\vspace{0.4cm}
\end{minipage}
}


\begin{abstract}
The absolute branching fractions for the decays $D^0 \to
K^-e ^+\nu _e$ and $D^0 \to \pi^-e^+\nu_e$ are determined
using $7584\pm 198 \pm 341$ singly tagged $\overline D^0$ 
sample from the data collected around 3.773 GeV
with the BES-II detector at the BEPC. In the system recoiling against
the singly tagged $\overline D^0$ meson, $104.0\pm 10.9$ events
for $D^0 \to K^-e ^+\nu _e$ and $9.0 \pm 3.6$ events for
$D^0 \to \pi^-e^+\nu_e$ decays are observed. Those yield the absolute
branching fractions to be 
$BF(D^0 \rightarrow K^-e^+\nu_e)=(3.82 \pm 0.40\pm 0.27)\%$ and
$BF(D^0 \rightarrow \pi^-e^+\nu_e)=(0.33 \pm 0.13\pm 0.03)\%$.
The vector form factors are determined to be
$|f^K_+(0)| = 0.78 \pm 0.04 \pm 0.03$ and
$|f^{\pi}_+(0)| = 0.73 \pm 0.14 \pm 0.06$. The ratio of the two form factors
is measured to be
$|f^{\pi}_+(0)/f^K_+(0)|= 0.93 \pm 0.19 \pm 0.07$.
\end{abstract}

\maketitle

\section{Introduction}

The semileptonic decays of the charmed mesons are theoretically
simplest to interpret because the effects of the weak and strong
interactions can be well separated. The decay amplitude is
proportional to the product of the Cabibbo-Kobayashi-Maskawa (CKM)
matrix element, which parametrizes the mixing between the quark
mass eigenstates and the weak eigenstates, and the form factor
describing the strong interaction between the final state quarks.
The differential decay rate for
$D^0\to K^-(\pi^- )e^+\nu_e$ process is described by
\begin{equation}
\frac {d\Gamma }{dq^2} = \frac {G_F^2}{24\pi ^3}|V_{cs(d)}|^2{\bf p}^3_{K(\pi)}
|f_+^{K(\pi )}(q^2)|^2 \label{rate},
\end{equation}
where $G_F$ is the Fermi coupling constant, $|V_{cs(d)}|$ is the CKM
matrix element and ${\bf p}_{K(\pi )}$ is the momentum of the
kaon(pion) in the rest frame of the $D^0$ meson. $f^{K(\pi )}_+(q^2)$ represents the 
vector form factor of the hadronic weak current depending on the square
of the four momentum transfer $q = p_D - p_{K(\pi )}$. 
In general theoretical treatment 
one common
form of the form factor is a single pole form and is expressed as
\begin{equation}
     f_+(q^2)=\frac{f_+(0)}{1-q^2/m^2_*} \label{pole}
\end{equation}
where $f_+(0)$ is the form factor evaluated at the four momentum 
transfer $q$ equal to zero, and the pole mass $m_*$ is the mass of 
the lowest-lying $Q \overline{q}^{\prime}$ meson.

   MARK III~\cite {mark}
previously made an absolute measurements of the branching fractions
for $D^0 \to K^-e^+\nu _e$ and $D^0 \to \pi ^-e^+\nu _e$ 
by analysing the data taken at the near $D\overline D$ threshold region. 
In this paper, we report the direct measurements
of the branching fractions for the
Cabibbo favored decay of 
$D^0 \to K^-e^+\nu_e$ (Throughout this paper, charged conjugation is implied.)
and the Cabibbo
suppressed decay of $D^0 \to \pi^-e^+\nu _e$ 
by analysing the data sample of integrated luminosity of $33~ \rm pb^{-1}$
collected at and around the center 
of mass energy of 3.773 GeV with the BES-II detector at the BEPC. 
Using the measured branching fractions, 
the well measured $|V_{cs}|$, $|V_{cd}|$
and the lifetime of the $D^0$ meson
quoted from PDG~\cite{pdg},
the vector form factors $|f^{\pi}_+(0)|$ and $|f^K_+(0)|$ 
are extracted, and their ratio
is determined directly.

\section{BES-II Detector}

BES-II is a conventional cylindrical magnetic detector that is
described in detail in Ref.~\cite{bes}.  A 12-layer Vertex Chamber
(VC) surrounding the beryllium beam pipe provides input to the event
trigger, as well as coordinate information.  A forty-layer main drift
chamber (MDC) located just outside the VC yields precise measurements
of charged particle trajectories with a solid angle coverage of $85\%$
of $4\pi$; it also provides ionization energy loss ($dE/dx$)
measurements which are used for particle identification.  Momentum
resolution of $1.7\%\sqrt{1+p^2}$ ($p$ in GeV/c) and $dE/dx$
resolution of $8.5\%$ for Bhabha scattering electrons are obtained for
the data taken at $\sqrt{s}=3.773$ GeV. An array of 48 scintillation
counters surrounding the MDC measures the time of flight (TOF) of
charged particles with a resolution of about 180 ps for electrons.
Outside the TOF, a 12 radiation length, lead-gas barrel shower counter
(BSC), operating in limited streamer mode, measures the energies of
electrons and photons over $80\%$ of the total solid angle with an
energy resolution of $\sigma_E/E=0.22/\sqrt{E}$ ($E$ in GeV) and spatial
resolutions of
$\sigma_{\phi}=7.9$ mrad and $\sigma_Z=2.3$ cm for
electrons. A solenoidal magnet outside the BSC provides a 0.4 T
magnetic field in the central tracking region of the detector. Three
double-layer muon counters instrument the magnet flux return, and serve
to identify muons of momentum greater than 500 MeV/c. They cover
$68\%$ of the total solid angle.

\section{Data Analysis}

At the center of mass energies around $3.773$ GeV, the
$\psi(3770)$ resonance is produced in electron-positron
($e^+e^-$) annihilation. The $\psi(3770)$
decays predominately into $D\overline D$ pairs. 
If one $D$ meson is fully reconstructed, 
the anti-$D$ meson must exist
in the system recoiling against the fully reconstructed $D$ meson (called
singly tagged $D$). Using the singly tagged $\overline D^0$ sample, the
decays of $D^0 \to K^-e^+\nu _e$ and $D^0 \to \pi ^-e^+\nu _e$
can be well selected in the recoiling system. Therefore, the absolute
branching fractions for these decays can be well measured.

\subsection{Event Selection}

The $\overline D^0$ meson is reconstructed in
non-leptonic decay modes of $K^+\pi^-$, $K^+\pi^-\pi^-\pi^+$,
$K^0\pi^+\pi^-$ and $K^+\pi^-\pi^0$.
Events which contain at least
two reconstructed charged tracks with good helix fits are selected.
In order to ensure
the well-measured 3-momentum vectors and the reliably charged particle
identification, the charged tracks used in the single tag analysis
are required to be within $|cos\theta|<$0.85 
where $\theta$ is the polar angle.
All tracks, save those from $K^0_S$ decays, must originate 
from the interaction region,
which require that the closest approach of a charged track
in $xy$ plane is less than 2.0 cm and
the $z$ position of the charged track is less than 20.0 cm.
Pions and kaons are identified by means of
TOF and $dE/dx$ measurements. Pion identification requires a consistency
with the pion hypothesis at a confidence level ($CL_{\pi}$) greater than
$0.1\%$.
In order to reduce misidentification, kaon candidate is
required to have a larger confidence level ($CL_{K}$) for a kaon hypothesis
than that for a pion hypothesis.
For electron identification,
the combined confidence level ($CL_{e}$), calculated for the $e$ hypothesis using
the $dE/dx$, TOF and BSC measurements, is required to be greater than
$0.1\%$, and the ratio
$CL_e/(CL_e + CL_{\pi} + CL_K)$ is required
to be greater than $0.8$.
The $\pi^0$ is reconstructed in the decay of $\pi^0 \rightarrow \gamma\gamma$. 
To select good photons from the decay
of $\pi^0$, the energy of a photon deposited in the BSC
is required to be greater than $0.07$ GeV, and the electromagnetic shower
is required to start in the first 5 readout layers. In order to reduce the
backgrounds the angle between the
photon and the nearest charged track is required to be greater than $22^{\circ}$
and the angle between the direction of the cluster development
and the direction of the photon emission to be less than $37^{\circ}$.
\subsection{Singly Tagged $\overline D^0$ Sample}
For each event, there may be several different charged track (or
charged and neutral track) combinations for each of
the four single tag modes.
Each combination is subject to center-of-mass energy constraint
kinematic fit and 
is required to have the fit probability $P(\chi^2)$ greater
than $0.1\%$. 
If more than one combination are satisfied $P(\chi^2)>0.1\%$,
the combination with the largest fit probability is retained.
For the single tag modes
with a neutral kaon and/or neutral pion,
one additional constraint kinematic fit
for the $K^0_S \rightarrow \pi^+\pi^-$ and/or
$\pi^0 \rightarrow \gamma\gamma$ hypothesis is performed, separately.

The resulting distributions in the fitted invariant masses 
of $Kn\pi$ ($n=1,2,3$) combinations,
which are calculated using the fitted momentum vectors from the kinematic
fit, are shown in Fig. 1.
The signals for the singly tagged $\overline D^0$ are clearly observed
in the fitted mass spectra.
A maximum likelihood fit 
to the mass spectrum with a Gaussian function for the $\overline D^0$
signal and a special
background function~\cite{fit} to describe backgrounds 
yields the number of the singly tagged $\overline D^0$ events 
for each of the four modes and the total number of $7584\pm 198 \pm 341$
reconstructed $\overline D^0$ mesons. The 
first error is statistical; the second error is systematic and is obtained by
varying the parameterization of the background.
\begin{figure}[hbt]
\includegraphics[width=8.5cm,height=8.5cm]
{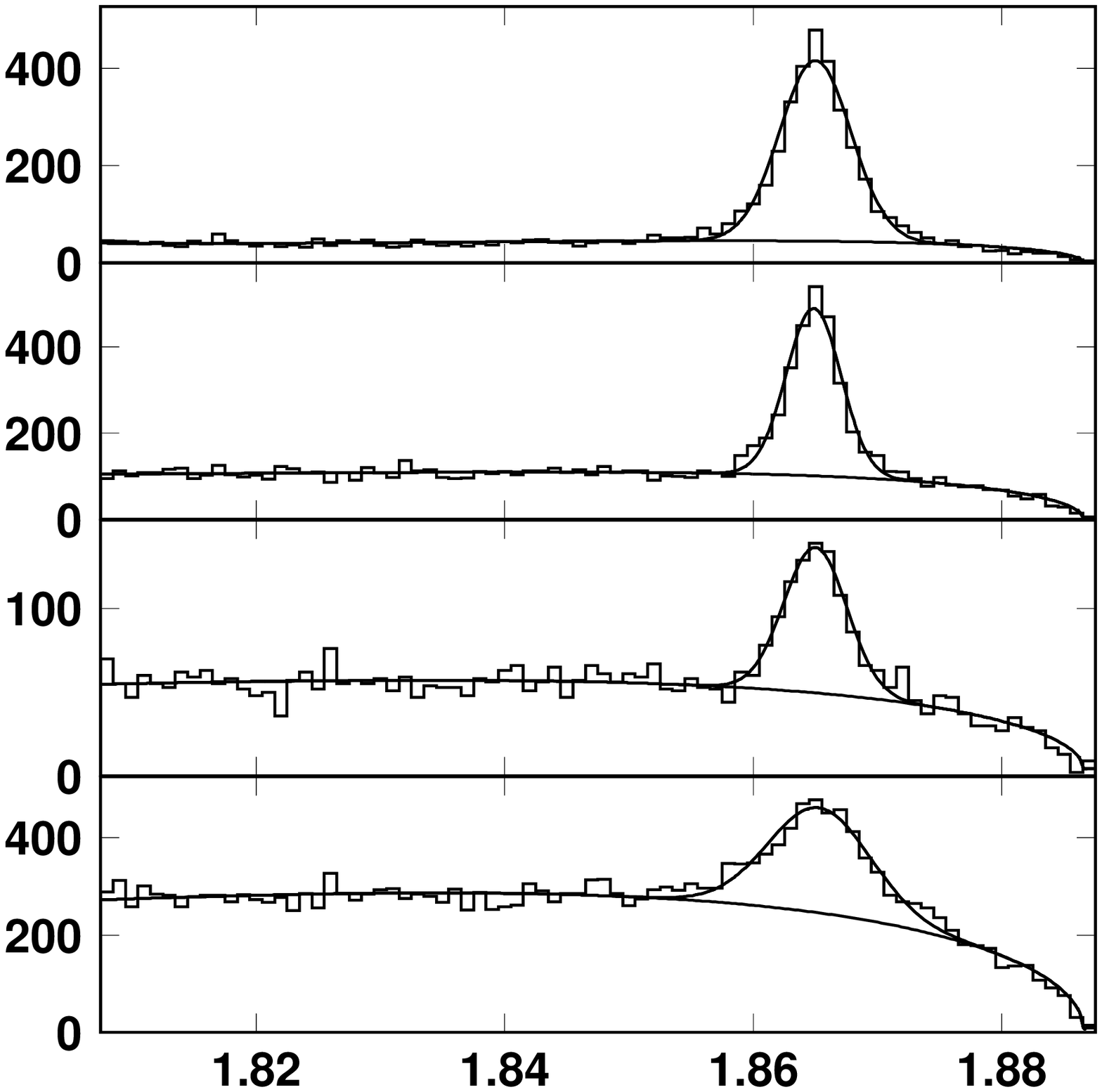}
\put(-190,215){(a)}
\put(-190,165){(b)}
\put(-190,112){(c)}
\put(-190,65){(d)}
\put(-150,-5){Invariant Mass (GeV/$c^2)$}
\put(-245,75){\rotatebox{90}{Events/(0.001 GeV/$c^2$)}}
\caption{Distributions of the fitted invariant masses of (a) $ K^+\pi^-$,
(b) $K^+\pi^-\pi^+\pi^-$, (c) $ K^0_S\pi^+\pi^-$ and
(d) $ K^+\pi^-\pi^0$ combinations.}
   \label{sgtag}
\end{figure}

\subsection{Candidates of $D^0 \rightarrow K^-e^+\nu_e$ and
$D^0 \rightarrow \pi^-e^+\nu_e$}

\begin{figure}[hbt]
\includegraphics[width=8.5cm,height=8cm]
{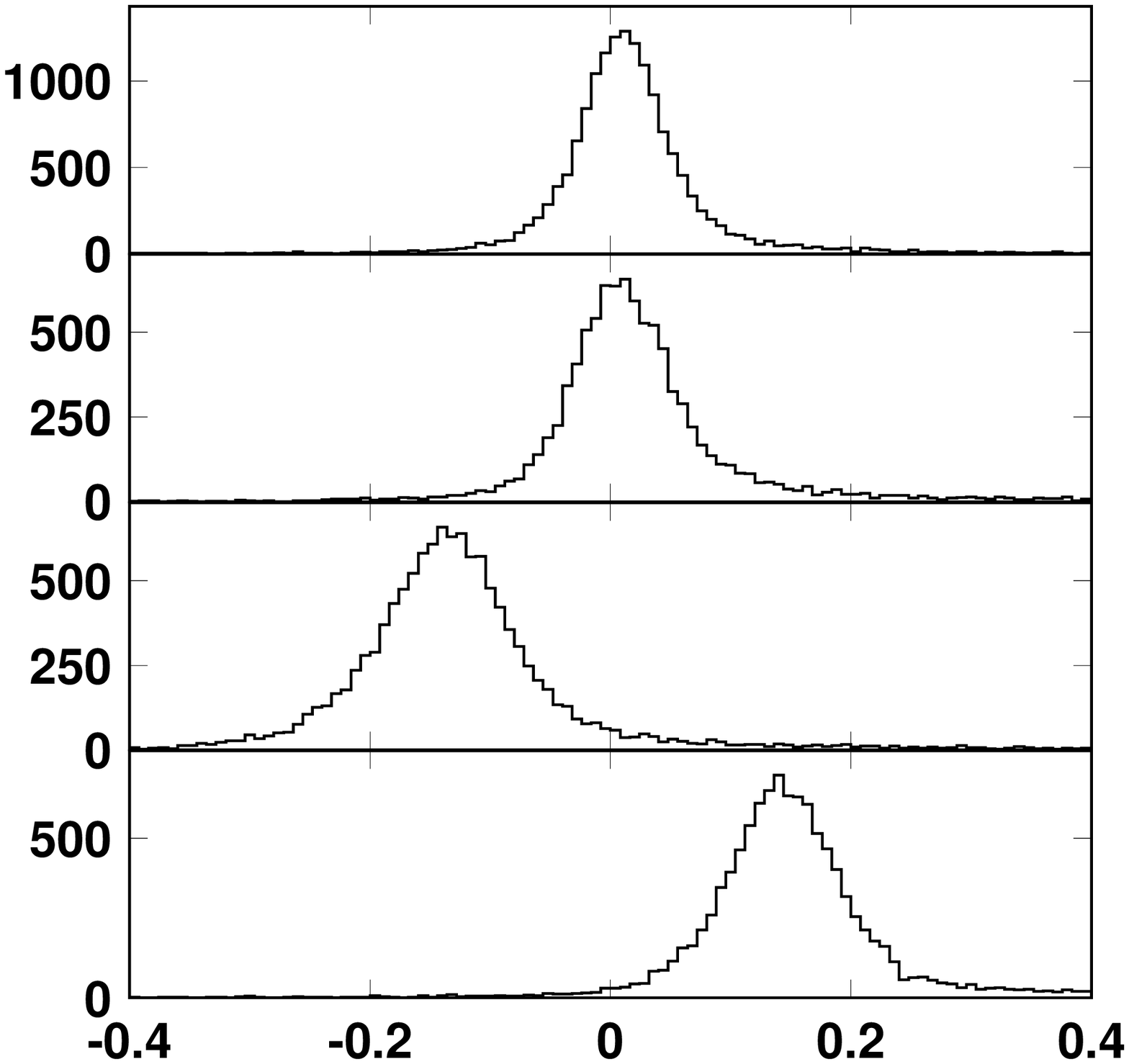}
\put(-185,200){\bf{(a)}}
\put(-185,150){\bf{(b)}}
\put(-185,105){\bf{(c)}}
\put(-185,60){\bf{(d)}}
\put(-130,-5){$U_{miss}$ (GeV)}
\put(-245,85){\rotatebox{90}{Events/(0.008 GeV)}}
\caption{Distribution of $U_{miss}$ calculated for the Monte Carlo events
of (a) $D^0\to K^-e^+\nu_e$, (b) $D^0\to\pi^-e^+\nu_e$,
(c) $D^0\rightarrow \pi^-e^+\nu$ by replacing pion mass with Kaon mass 
and
(d) $D^0 \to K^-e^+\nu_e$ by replacing kaon mass with pion mass.
}\label{umiss2}
\end{figure}

Candidate events $D^0\to K^-e^+\nu _e$ and $D^0\to \pi ^-e^+\nu _e$ are
selected from the surviving tracks in the system recoiling against the tagged 
$\overline D^0$. To select the $D^0 \to K^-e^+\nu_e$ and 
$D^0 \to \pi ^-e^+\nu_e$ events, it is required that there are only 
two oppositely charged tracks, 
one of which is identified as an electron and 
the other as a kaon or pion. 
The  neutrino is undetected, therefore the kinematic
quantity  $U_{miss}\equiv E_{miss}-p_{miss}$ is used
to obtain the information about the missing neutrino, where $E_{miss}$ and 
$p_{miss}$ are the total energy
and the momentum of all missing particles respectively. 
Fig.~\ref{umiss2}(a) and Fig.~\ref{umiss2}(b) 
show the $U_{miss}$ distributions for
the Monte Carlo $D^0 \to K^-e^+\nu _e$ and $D^0 \to \pi^-e^+\nu_e$ events
respectively. The candidate events are required to satisfy
the requirement $|U_{miss}| < 3\sigma_{U_{miss}}$, where the
$\sigma_{U_{miss}}$ is
the standard deviation of the $U_{miss}$ distribution.

The branching fraction of the Cabibbo favored decay
$D^0 \to K^-e^+\nu_e$ is much larger than that of the Cabibbo 
suppressed decay $D^0 \to \pi ^-e^+\nu _e$. The kaon can be 
misidentified as pion and therefore the process $D^0 \to 
K^-e^+\nu_e$ in the recoil side can be misclassified as 
$D^0 \to \pi^-e^+\nu_e$. Monte Carlo study shows that this 
decay process is the main contamination to 
the selected sample of $D^0 \to \pi^-e^+\nu_e$ process.
In order to correctly select the events
$D^0 \to \pi^-e^+\nu_e$ and suppress misidentification from
$D^0 \to K^-e^+\nu_e$, the quantity $U_{\pi-as-K}$ is 
calculated by 
replacing pion mass with the kaon mass and 
$|U_{miss}|<|U_{\pi-as-K}|$ is required.
Fig.~\ref{umiss2}(c) shows the $U_{miss}$ calculated by replacing pion
mass with kaon mass for the Monte Carlo events
of $D^0\rightarrow \pi^-e^+\nu$, while
Fig.~\ref{umiss2}(d) shows the distributions of $U_{miss}$ calculated
by replacing kaon mass with pion mass
for the Monte Carlo events of $D^0 \to K^-e^+\nu_e$. 
The quantity $U_{miss}$ is expected to be closer to zero for the 
correct particle mass assignment.
The decays such as $D^0 \rightarrow K^-\pi^0 e^+(\mu^+) \nu_e (\nu_{\mu})$
are suppressed by rejecting the events with extra isolated
photons which are not
used in the reconstruction of the singly tagged $\overline D^0$.
The isolated photon should have its energy greater than 0.1 GeV and
satisfy photon selection criteria as mentioned earlier.

\begin{figure}[hbt]
\includegraphics[width=8.5cm,height=7.5cm]
{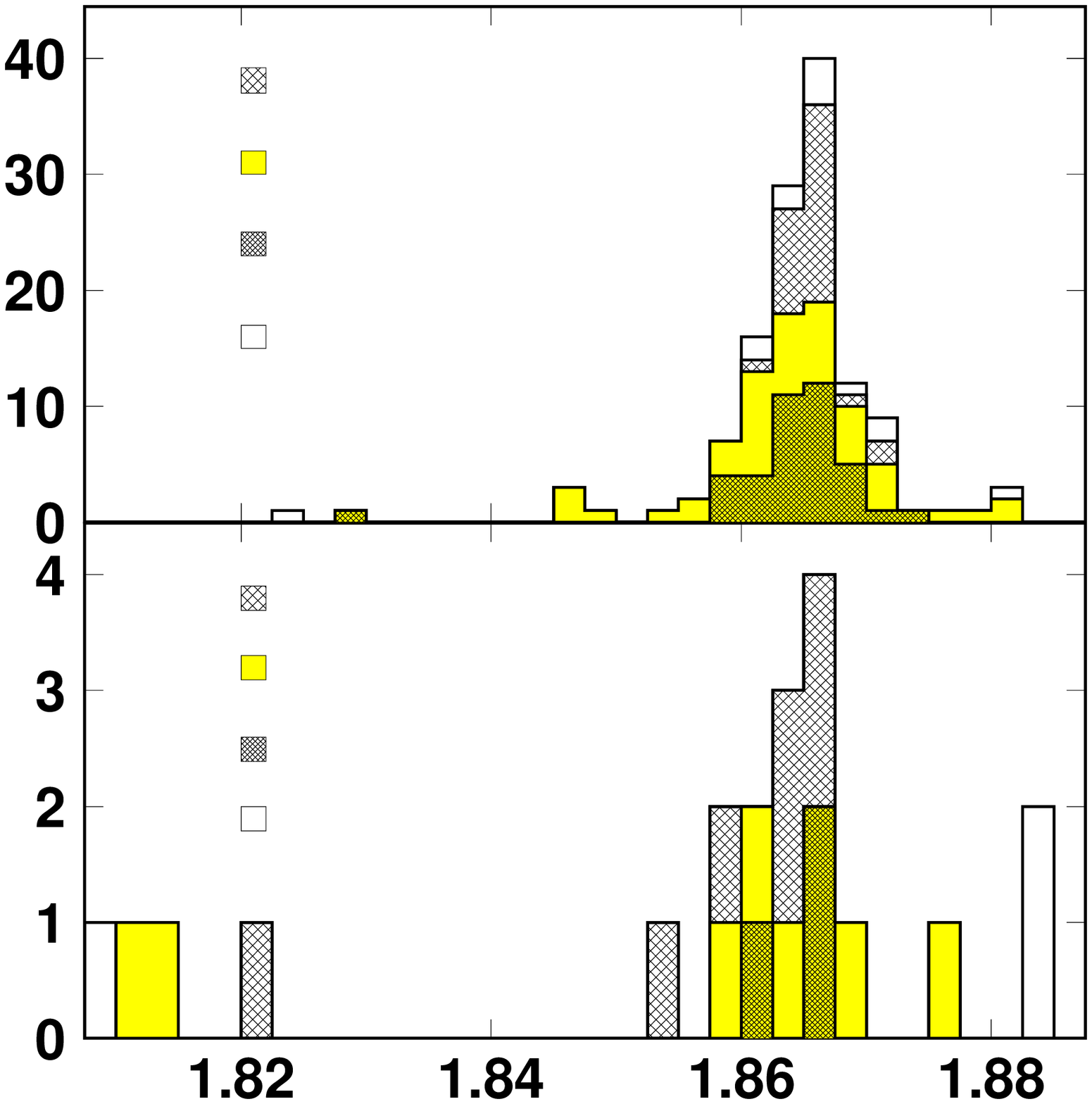}
\put(-190,180){\bf{(a)}}
\put(-190,90){\bf{(b)}}
\put(-160,187){\tiny {$\overline D^0 \rightarrow K^+\pi^-\pi^-\pi^+$}}
\put(-160,175){\tiny {$\overline D^0 \rightarrow K^+\pi^-\pi^0$}}
\put(-160,160){\tiny {$\overline D^0 \rightarrow K^+\pi^-$}}
\put(-160,145){\tiny {$\overline D^0 \rightarrow K^0\pi^+\pi^-$}}
\put(-160,99){\tiny {$\overline D^0 \rightarrow K^+\pi^-\pi^-\pi^+$}}
\put(-160,87){\tiny {$\overline D^0 \rightarrow K^+\pi^-\pi^0$}}
\put(-160,74){\tiny {$\overline D^0 \rightarrow K^+\pi^-$}}
\put(-160,62){\tiny {$\overline D^0 \rightarrow K^0\pi^+\pi^-$}}
\put(-150,-7){Invariant Mass (GeV/$c^2)$}
\put(-245,85){\rotatebox{90}{Events/(0.0025 GeV/$c^2$)}}
\caption{Distributions of the fitted invariant masses of $Kn\pi$
combinations for the events in which (a) the
          $D^0 \rightarrow K^-e^+\nu_e$ and 
          (b) the $D^0 \rightarrow \pi^-e^+\nu_e$
          candidate events are observed in the system recoiling against the
tagged $\overline D^0$.}
  \label{dbtag}
\end{figure}

Fig.~\ref{dbtag}(a) and Fig.~\ref{dbtag}(b) show the distributions of
the fitted invariant masses of the $Kn\pi$ combinations
for the events in which the  $D^0 \to K^-e^+\nu_e$ and 
$D^0 \to \pi^-e^+\nu_e$ candidate events are observed 
in the system recoiling against the singly tagged $\overline D^0$.
In the Fig.~\ref{dbtag}(a), there are 118 events
in the $\pm 3\sigma$ signal regions,
while there are 10 events in the 
outside of the signal regions.
By assuming that the background distribution is flat, 
$3.8 \pm 1.3$ background events are estimated in the signal region. 
After subtracting the number of background events in the signal region, 
$114.2 \pm 10.9$ candidate events are retained.
A similar analysis of the events in Fig.~\ref{dbtag}(b) gives
that there are $11.0 \pm 3.6$ candidate events 
after subtracting the number of background events in the signal region.

\begin{figure}[hbt]
\includegraphics[width=9.0cm,height=8cm]
{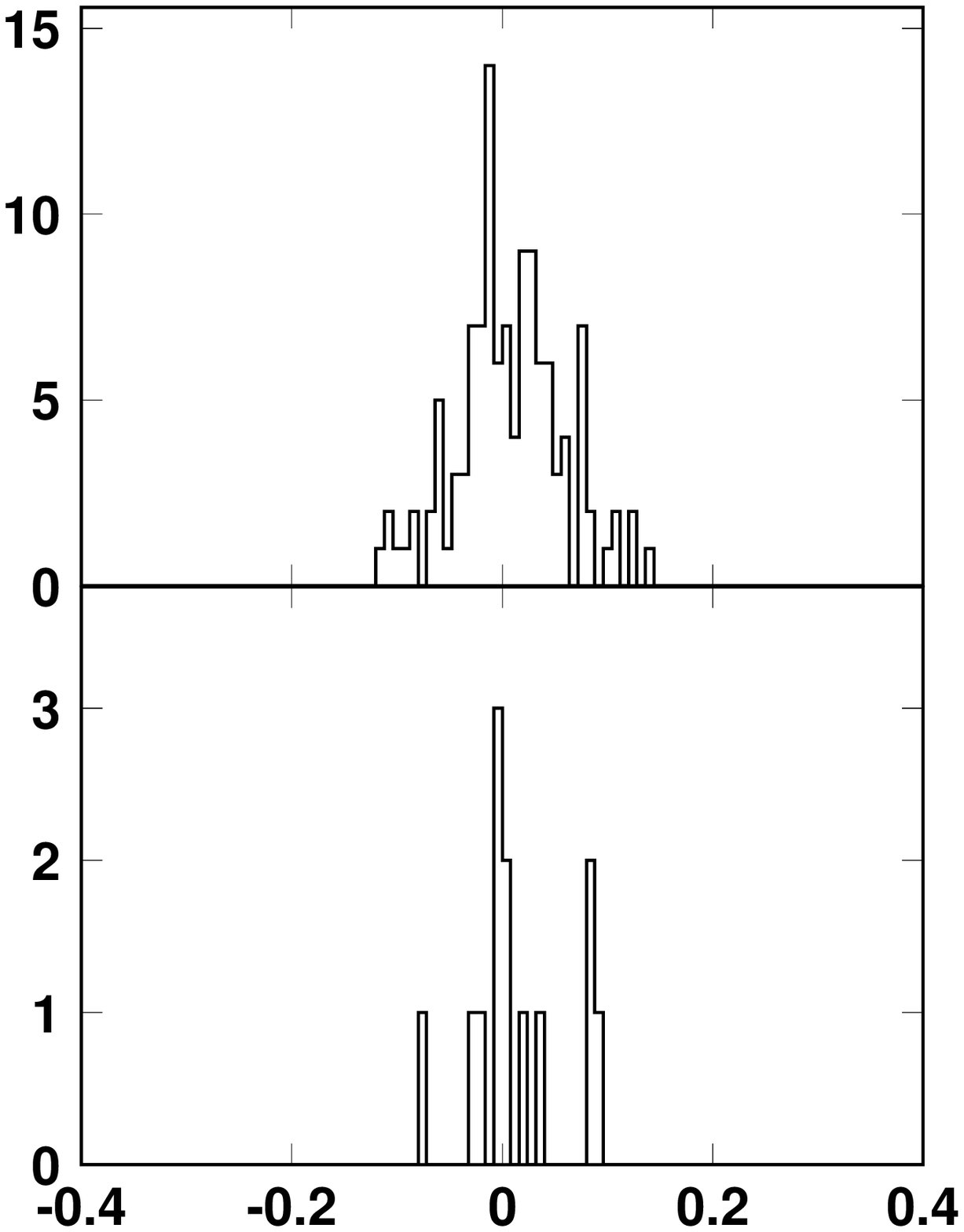}
\put(-185,200){\bf{(a)}}
\put(-185,105){\bf{(b)}}
\put(-130,-5){$U_{miss}$ (GeV)}
\put(-245,85){\rotatebox{90}{Events/(0.008 GeV)}}
\caption{Distribution of $U_{miss}$ calculated for the selected candidate events
of (a) $D^0\to K^-e^+\nu_e$ and (b) $D^0\to\pi^-e^+\nu_e$.
}\label{umiss_data}
\end{figure}

\begin{figure}[hbt]
\includegraphics[width=9.0cm,height=8cm]
{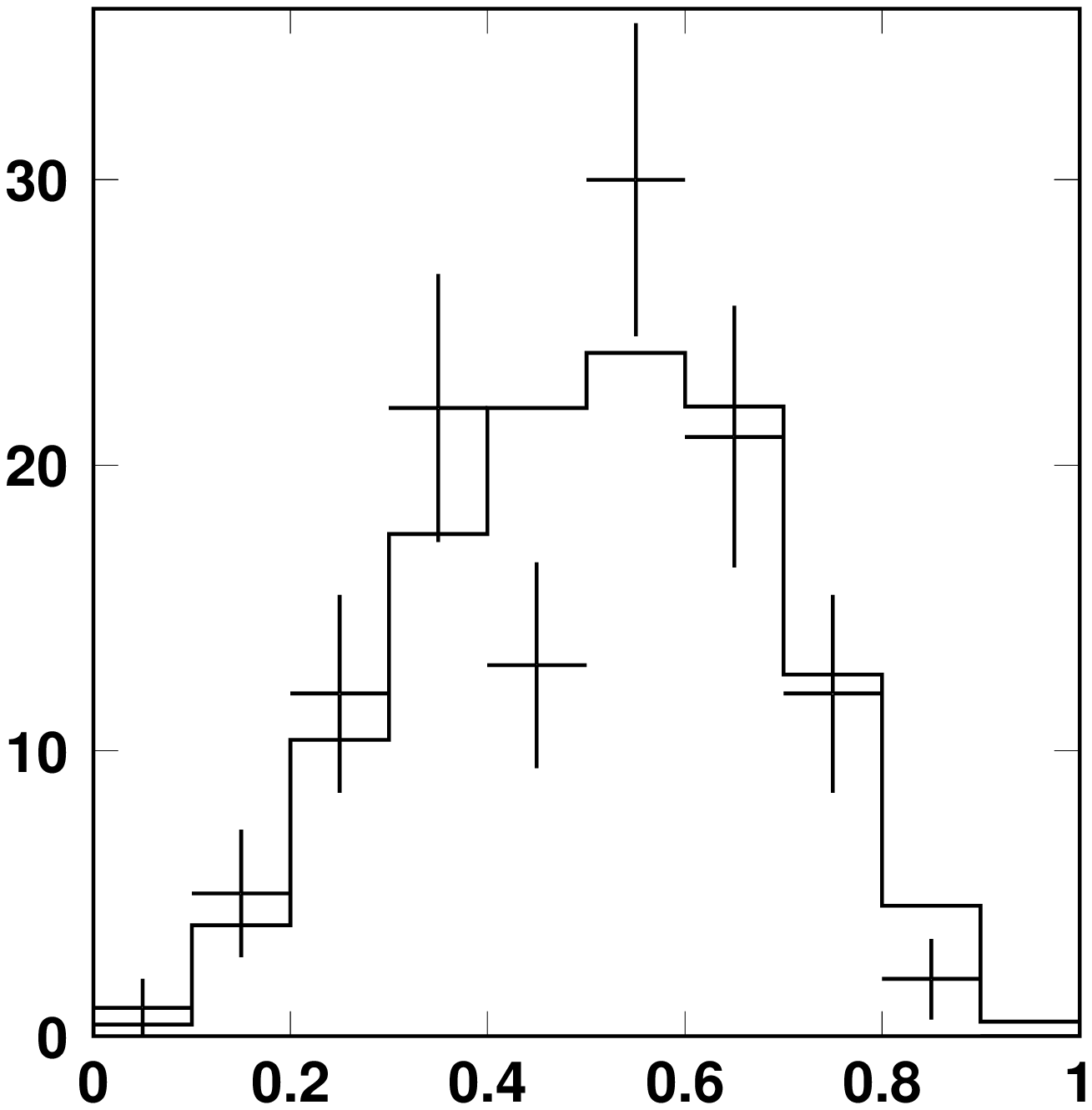}
\put(-160,-2){Momentum of electron~~~~ (GeV/$c$)}
\put(-245,85){\rotatebox{90}{Events/(0.1 GeV)}}
\caption{Distribution of the momentum of the electrons from the
selected candidate events of $D^0\to K^-e^+\nu_e$,
where the error bars are for the events from the
data and the histogram is for the events of
$D^0 \rightarrow K^-e^+\nu_e$ from Monte Carlo sample.
}\label{momentum_e}
\end{figure}

\noindent
Fig.~\ref{umiss_data}(a) and Fig.~\ref{umiss_data}(b) show
distributions of the $U_{miss}$ calculated for the selected events of
$D^0 \rightarrow K^-e^+\nu$ and $D^0 \rightarrow \pi^-e^+\nu$,
respectively.
Fig.~\ref{momentum_e} shows distribution of the momentum of the
electrons from the selected candidate events of
$D^0 \rightarrow K^-e^+\nu_e$, where the error bars are for the events from
the data and the histogram is for the events of $D^0 \rightarrow
K^-e^+\nu_e$
from Monte Carlo sample.

\subsection{Background Subtraction}

There are still some background contaminations in the observed
candidate events due to other semileptonic or 
hadronic decays. These background events must be subtracted 
from the candidate events. 
The numbers of background are estimated by analysing the 
Monte Carlo sample which is 13 times larger than the data. 
The Monte Carlo events are generated as
$e^+e^- \rightarrow D \overline D$ and the $D$ and $\overline D$
mesons are set to decay to all possible final states
according to the decay modes and branching fractions 
quoted from PDG~\cite{pdg} except the two decay modes under study.
The number of events satisfying the selection criteria is then renormalized
to the corresponding data set.
Totally $10.2 \pm 1.0$ and $2.0 \pm 0.5$ background
events are obtained for $D^0 \rightarrow K^-e^+\nu _e$
and $D^0 \rightarrow \pi ^-e^+\nu _e$, respectively.
After subtracting these numbers of background events,
$104.0\pm 10.9$ and $9.0 \pm 3.6$ signal events
for $D^0 \rightarrow K^-e^+\nu _e$ and 
$D^0 \rightarrow \pi ^-e^+\nu _e$ decays are retained.

\section{Results}

\subsection{Monte Carlo Efficiency}
The efficiencies for reconstruction of the semileptonic decay
events of $D^0 \to K^-e^+\nu _e$ and
$D^0 \to \pi ^-e^+\nu _e$ are estimated
by Monte Carlo simulation. A detailed Monte Carlo study shows
that the efficiencies are
$\epsilon_{K^-e^+\nu _e}=(35.89 \pm 0.25)\%$ and
$\epsilon_{\pi^-e^+\nu_e}=(36.02 \pm 0.25)\%$,
where the errors are statistical.

\subsection{Branching Fractions}

The measured branching fractions are obtained by dividing 
the observed numbers of the semileptonic decay events
$N(D^0 \rightarrow K^-(\pi^-) e^+ \nu_e)$
by the number of the singly tagged $\overline D^0$
meson $N_{\overline D^0_{tag}}$ and the reconstruction efficiencies
$\epsilon_{K^-e^+\nu_e(\pi^-e^+\nu_e)}$,
\begin{equation}
Br(D^0\to K^-(\pi^- )e^+\nu _e)=
\frac{ N(D^0 \rightarrow K^-(\pi^-) e^+ \nu_e) }
{ \epsilon_{K^-e^+\nu_e(\pi^-e^+\nu_e)} \times N_{\overline D^0_{tag}}}.
\end{equation}
Inserting these numbers
into the equation (3), 
the branching fractions
for $D^0\to K^-e^+\nu_e$ and $D^0\to \pi^-e^+\nu_e$ decays
are obtained to be
$$BF(D^0 \to K^-e^+\nu_e)=(3.82\pm 0.40 \pm 0.27)\%$$
\noindent and
$$BF(D^0 \to \pi^- e^+\nu_e)=(0.33\pm 0.13\pm 0.03)\%,$$
where the first errors are statistical and the second systematic.
The systematic uncertainties in the measured branching 
fractions arise from the uncertainties of
particle identification $(1.1\%)$, tracking efficiency 
($2.0\%$ per track), photon reconstruction ($2.0\%$), 
$U_{miss}$ selection ($0.6\%$ for $D^0 \to K^-e^+\nu_e$, 
$1.2\%$ for $D^0 \to \pi^-e^+\nu_e$), 
the number of the singly tagged $\overline D^0$
($4.8\%$), background subtraction 
($2.3\%$ for $D^0 \to K^-e^+\nu_e$, 
$5.6\%$ for $D^0 \to \pi^-e^+\nu_e$) and 
Monte Carlo statistics ($0.8\%$).
These uncertainties are added in quadrature to obtained the total systematic
errors, which are $7.1\%$ and $8.8\%$ for $D^0 \to K^-e^+\nu_e$ and
$D^0 \to \pi^-e^+\nu_e$, respectively.
\subsection{Form Factors $|f^K_+(0)|$ and $|f^\pi_+(0)|$}
The decay width~\cite{ffk}\cite{ffpi} of the semileptonic decay processes 
can be derived from the equation~(\ref{rate})
by substituting the single pole form of the
form factor as given in equation~(\ref{pole})
for $|f_+^{K(\pi)}(q^2)|$ in the equation (1).
The relations between the decay widths and the form factors are
\begin{equation}
\Gamma(D^0 \to K^-e^+\nu_e) = 1.53\;|V_{cs}|^2 
|f^K_+(0)|^2 \times 10^{11} s^{-1},
\end{equation}

\begin{equation}
\Gamma(D^0 \to \pi^-e^+\nu_e) = 3.01\; |V_{cd}|^2 
|f^{\pi}_+(0)|^2\times 10^{11} s^{-1}.
\end{equation}
The form factors $|f^K_+(0)|$ and $|f^{\pi}_+(0)|$
can be extracted by using the measured values
of the branching fractions and the lifetime
of the $D^0$ meson. Inserting the values of $|V_{cs}| = 0.996 \pm 0.013$,
$|V_{cd}| = 0.224 \pm 0.016$ and the lifetime 
$\tau_{D^0} = (411.7 \pm 2.7) \times 10^{-15}~s$ into equation (4) and
(5), the form factors are obtained to be
$$|f^K_+(0)| = 0.78 \pm 0.04 \pm 0.03,$$
$$|f^{\pi}_+(0)| = 0.73 \pm 0.14 \pm 0.06,$$
\noindent
where the first errors are statistical and the
second are systematic errors which arise from the systematic uncertainties
in the measured values of the branching fractions, the uncertainties in
the values of $|V_{cs}|$, $|V_{cd}|$ and $\tau_{D^0}$.
The values of the form factors are compared with that
predicted by various theoretical models and enumerated in Table I.
\begin{table}[hbt]
\caption{{\small Form factor.}}
\begin{center}
\begin{tabular}{ccc} \hline\hline
           & $|f^K_+(0)|$ & $|f^{\pi}_+(0)|$\\
\hline
QCDSR~\cite{qcd}& $0.78 \pm 0.11$ & $0.65 \pm 0.11$\\ 
LQCD1~\cite{lattice1}& $0.71 \pm 0.03^{+0.00}_{-0.07}$&
                             $0.64 \pm 0.05^{+0.00}_{-0.07}$\\ 
LQCD2~\cite{lattice2}& $0.66\pm 0.04^{+0.01}_{-0.00}$&
                             $0.57\pm 0.06^{+0.01}_{-0.00}$\\                             
BES& $0.78 \pm 0.04 \pm 0.03$ & $0.73 \pm 0.14 \pm 0.06$\\ 
\hline \hline
\end{tabular}
\end{center}
\end{table}
The ratio of the two form factors can be obtained from the equations
(3), (4) and (5),
\begin{equation}
\frac{|f^{\pi}_+(0)|} {|f^{K}_+(0)|} = 0.71\frac{|V_{cs}|}{|V_{cd}|}
       \sqrt{
 \frac{N(D^0\rightarrow \pi^-e^+\nu_e)\epsilon_{K^-e^+\nu_e} }
      {N(D^0\rightarrow   K^-e^+\nu_e)\epsilon_{\pi^-e^+\nu_e}} } ,
\end{equation}
Inserting the $|V_{cd}|$, $|V_{cs}|$, the numbers of the signal
events and the efficiencies into the equation (6), the value of the ratio
is obtained. It is
$$|f^{\pi}_+(0)/f^K_+(0)|= 0.93 \pm 0.19 \pm 0.07,$$
\noindent
where the first error is statistical and
the second systematic which arises from the systematic uncertainties
in the measured values of the branching fractions and the uncertainties in
the values of $|V_{cs}|$ and $|V_{cd}|$.
This result is consistent with theoretical predictions, which range
from 0.7 to 1.4~\cite{richman}.

\subsection{CKM Matrix Elements  $|V_{cs}|$ and $|V_{cd}|$}
Reversing the argument that presented in the previous section, we obtain
the measured values of the CKM matrix elements $|V_{cs}|$ and $|V_{cd}|$
using the predicted form factors as shown in Table I. 
The results are listed in Table II. As a comparison, the values
of the CKM matrix elements quoted from PDG~\cite{pdg}
are also listed in the Table II.
\begin{table}[hbt]
\caption{{\small CKM matrix element.}}
\begin{center}
\begin{tabular}{ccc} \hline\hline
$|V_{cs}|$ & $|V_{cd}|$ & $f_+^{K(\pi)}(0)$ input \\
\hline
$0.998 \pm 0.052 \pm 0.145$ &
                    $0.251 \pm 0.049\pm 0.044$ &  QCDSR~\cite{qcd}\\
$1.097 \pm 0.057^{+0.061}_{-0.124}$& 
     $0.255 \pm 0.050^{+0.023}_{-0.036}$ & LQCD1~\cite{lattice1}\\
$1.180 \pm 0.062^{+0.085}_{-0.083}$& 
     $0.286 \pm 0.056^{+0.033}_{-0.033}$ & LQCD2~\cite{lattice2}\\
\hline
$0.996 \pm 0.013$ & $0.224 \pm 0.016$ & PDG\\
\hline \hline
\end{tabular}
\end{center}
\end{table}

Finally, Table III gives the comparison of the ratio of the 
CKM matrix elements with that obtained by the MARKIII, in which
the ratio of the form factors is taken to be unity.
\begin{table}[hbt]
\caption{{\small The ratio of the CKM matrix elements.}}
\begin{center}
\begin{tabular}{ccc} \hline\hline
           & BES & MARKIII~\cite{mark}\\
\hline
$|V_{cd}/V_{cs}|^2$ & $0.043 \pm 0.017 \pm 0.003$ &
                                    $0.057^{+0.038}_{-0.015} \pm 0.005$ \\
\hline \hline
\end{tabular}
\end{center}
\end{table}

\section{Summary}

In summary, by analysing the data collected at and around 3.773 GeV 
with the BES-II detector at the BEPC, the branching fractions for the decay
of $D^0 \rightarrow K^-e^+\nu_e$ and  $D^0 \rightarrow \pi^-e^+\nu_e$
have been measured. From a total of $7584\pm 198 \pm 341$ 
singly tagged $\overline D^0$ sample, 
$104.0\pm 10.9$ $D^0 \to K^-e^+\nu _e$ and
$9.0 \pm 3.6$  $D^0 \to \pi ^-e^+\nu _e$ signal events
are observed in the system recoiling against the $\overline D^0$
tags. Those yield the decay branching fractions to be
$BF(D^0 \to K^-e^+\nu_e)=(3.82\pm 0.40\pm 0.27)\%$ and
$BF(D^0 \to \pi^-e^+\nu_e)=(0.33\pm 0.13\pm 0.03)\%$.
Using the values of the CKM matrix elements quoted from PDG~\cite{pdg},
the form factors $|f^K_+(0)|$ and
$|f^K_+(0)|$ are determined to be 
$|f^K_+(0)| = 0.78 \pm 0.04 \pm 0.03$,
$|f^{\pi}_+(0)| = 0.73 \pm 0.14 \pm 0.06$ and
the ratio of the two form factor to be
$|f^{\pi}_+(0)/f^K_+(0)|= 0.93 \pm 0.19 \pm 0.07.$
In addition, using the form factors predicted by QCDSR and
LQCD calculations, the CKM matrix elements $|V_{cs}|$ and $|V_{cd}|$
are also determined.

\vspace{5mm}

\begin{center}
{\small {\bf ACKNOWLEDGEMENT}}
\end{center}
\par
\vspace{0.4cm}

   The BES collaboration thanks the staff of BEPC for their hard efforts.
This work is supported in part by the National Natural Science Foundation
of China under contracts
Nos. 19991480,10225524,10225525, the Chinese Academy
of Sciences under contract No. KJ 95T-03, the 100 Talents Program of CAS
under Contract Nos. U-11, U-24, U-25, and the Knowledge Innovation Project
of CAS under Contract Nos. U-602, U-34(IHEP); by the National Natural Science
Foundation of China under Contract No.10175060(USTC),and
No.10225522(Tsinghua University).

\end{document}